# FSS++ Workshop Report:
# Handling Uncertainty for Data Quality Management

*Challenges from Transport and Supply Chain Management*

Anna Wilbik  (editor)
School of Industrial Engineering
Eindhoven University of Technology
a.m.wilbik@tue.nl

## 1. Introduction

Currently more and more data is collected. The volume of business data worldwide, across all companies, is estimated to double every 1.2 years [1]. Availability of data and data analytics methods opens new possibilities in many business domains. The logistics and supply chain domain is an important case among these, in which the proper use of data can facilitate a spectrum of improvements in strategic and tactical planning, as well as in operations management.

Applications for (big) data analytics have been found among others in demand and production planning, inventory management, and transportation logistics [5,7]. Future applications can address agility of supply chains, better use of Internet of Things (IoT) technology and sustainability issues. Companies expect to benefit from big data business analytics in logistics and supply chain operations by improving the visibility, flexibility, and integration of global supply chains and logistics processes, effectively manage demand volatility, and handle cost fluctuations [6]. For instance, small business organizations can prevent bottlenecks throughout the supply chain by analyzing real-time data. Companies can gain transparency needed to control inventory fluctuations - either in the procured goods used to manufacture products or in the finished goods being sold and distributed to consumers [4]. Traditional supply chain data can be extended with additional data such as weather data, data on events, like outbreaks of flu, and data collected from news streams [8]. This additional data can generate insights for the short term, such as how operations will be affected in the current week [2].

Big data analytics offer huge opportunities to the logistics and supply chain domains, however there are also threats. Companies are trying to capture and store everything, without first establishing the business utility of data [3]. This may result in increased processing time and misleading insights delivered from irrelevant data. Another threat is related to data quality. Many companies are struggling with having and keeping the data accurate [9].

A few numbers below illustrate the magnitude of this problem. IBM has estimated the yearly cost of poor quality data, in the US alone, to be $3.1 trillion [10]. Moreover, it is believed that knowledge workers waste up to 50% of time hunting for data, identifying and correcting errors, and seeking confirmatory sources for data they do not trust [11]. Data scientists spend 60% of their time on cleaning and organizing data on top of 19% for data collection [12]. There are examples where a bill of materials is only 10% accurate [9].

This report describes the results of the eSCF Awareness Workshop on Handling Uncertainty for Data Quality Management - Challenges from Transport and Supply Chain Management that was held on June 5, 2018 in Heeze, The Netherlands. The goal of this workshop was to create and enhance awareness into data quality management issues that are encountered in practice, for business organizations that aim to integrate a data-analytical mind set into their operations.

This report is structured as follows. The next section describes most common data quality dimensions, which forms the background for the workshop. Section 3 presents the set-up for the workshop, while Section 4 describes the results of the workshop. This report is finished with concluding remarks based on these results.

## 2. Background

Data quality problems have been investigated for a long time. Researchers have identified many dimensions and criteria as well as several classifications [13,14,15]. From those the following four dimensions are the most important ones [16]:

- completeness,
- accuracy,
- consistency,
- timeliness.

We explain them briefly, as they can have different meanings. Completeness is very often understood as the degree to which the data (or attributes of data) that were intended for collection were collected in reality. This measure should take into account inapplicability, i.e., in some cases empty values are expected (e.g., in a person database, a single person has an empty value for the *spouse* attribute). However completeness sometimes is understood more as coverage, i.e., to what degree the data set in terms of collected objects is complete. The third aspect of completeness is related to the notion of density, i.e. whether enough attributes are being collected [13]. Accuracy can have also several interpretations. It can describe whether the data is correct, reliable, precise or certified to be free of errors. So in other words, this dimension measures the quantity of both incorrect data and inaccurate data [14]. Consistency is often meant as integrity. It can be defined as a degree to which data managed in a system satisfy specified constraints or business rules [17]. Last but not least, timeliness denotes the degree to which provided data is up-to-date [17].

## 3. Workshop setup

The aim of the workshop was to discuss the impact and frequency of selected data quality problems as well as possible solutions for the most important ones. 23 people participated in the workshop, including two company representatives, 14 researchers and seven PhD students. The discussion was scoped towards five specific data quality problems:

- missing data,
- incorrect data,
- inaccurate data,
- poorly defined data,
- contradictory data.

Those problems are related to sub-dimensions of three of those most important dimensions of data quality as we discussed before: completeness, accuracy and consistency of data. Timeliness of data is considered to be out-of-scope for this workshop, as we did not focus on real-time data characteristics.

The workshop consisted of two parts. In the first part, participants were asked to discuss the frequency and impact of the discussed data quality problems with the help of the *data quality matrix*. Participants were working in small groups and the findings of group work were shared and discussed with all participants.

The *data quality matrix* is a simple tool that aids in setting up a structured discussion about the types of data quality problems, the frequency that a type of problem occurs, and the impact of occurring problems. An empty matrix is shown in Figure 1. The horizontal dimension of the matrix represents the five selected data quality problems, the vertical dimension represents frequency and impact. Problems that occur *never* or *rarely* and have a *very low impact* or *low impact* are obviously of a much lesser concern to industrial practice than problems that occur *always* or *often* and have a *very high impact* or *high impact*. Consequently, this matrix can be used to set priorities for data quality management improvements in practice.

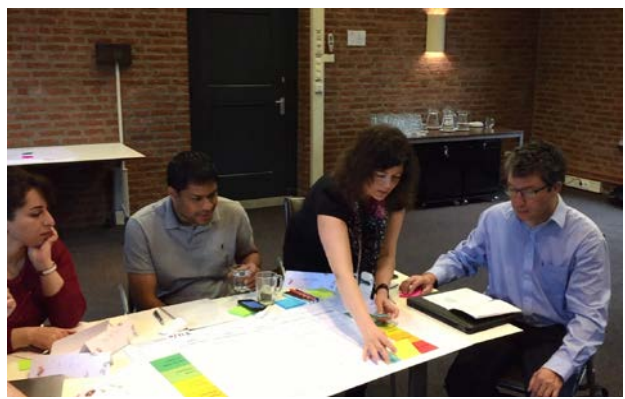

*Fig. 1. Data quality matrix*

In the second part of the workshop, participants were discussing and brainstorming regarding different possibilities and solutions towards problems that were identified in the first part. Like in the first part of the workshop, initially participants were working in smaller groups, and later the most important ideas were exchanged and discussed with all participants.

*Fig. 2. Participants filling the data quality matrix*

## 4. Results and discussion

Below we present the main results from the discussions during the workshop. First we discuss the results of the application of the data quality matrix.

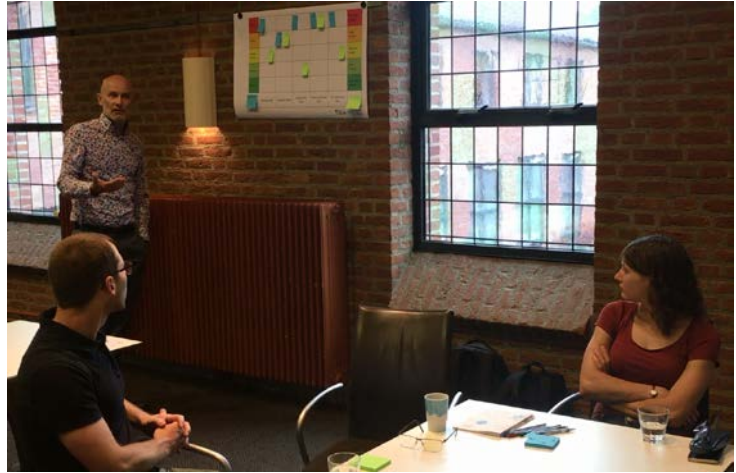

*Fig. 3. Participants discussing the results*

All the participants agreed that missing data is a problem that occurs almost always, however there was no consensus reached regarding the impact of this problem: the impact was assessed to range from very high to very low.

The problem of incorrect data is believed to be observed often or sometimes. The impact of this problem was believed to be between very high and low with also quite diverse opinions.

All participants agreed that the problem of inaccurate data is almost always present, however the impact of this problem was not considered to be high (very low to medium).

The analysis of the problem of poorly defined data showed an interesting result. One group containing business representatives found this problem omnipresent and with huge impact. To the contrary, the other groups consisting of researchers concluded that this problem is present sometimes or rarely and has limited impact. This may indicate a misalignment of industry requirements and research efforts.

Regarding the problem of contradictory data no agreement was reached with respect to the frequency or impact.

In the second part of the workshop, participants were discussing how to overcome the identified problems. The first observation concerns the data collection and can influence all discussed problems. Proper data collection can and will prevent many data quality problems and save much time, effort and money during cleansing the data. The purpose of the data collection should be known upfront. A strategy "collect the data and then we will see" often does not work. Moreover the quality of data should be monitored all the time, during the data collection, before any analysis starts. Information about data quality can be stored in data quality marts, which can enable easy and regular reporting. Those reports should be made available to all employees dealing with the data, so that they can see the effect of their actions and feel responsible for the data collection. Employees should also know how good the data is to estimate the reliability and confidence of models built on the basis of this data. Automation of the data collection processes can improve data quality.

A problem that is connected to these general remarks is poorly defined data. In order to prevent problems with poorly defined data, the company can create a data glossary. This data glossary is a dictionary type document that defines and explains all the data which are collected by the company. Creating such data glossary, or in a simpler form a meta-data structure, is a tedious work and it is hardly ever done, but can solve many data quality problems.

A very common data quality problem is missing data. The standard solutions include removing observations or removing variables from the data set or imputing the missing values. However not so much attention is paid to the fact that sometimes missing data carry the information by itself. So far this aspect has not been investigated in research and is not used in current data science practice. Another issue with missing data is that not all variables are equally important: one variable may me more important than another. This should be taken into consideration during cleansing process or evaluating data quality.

Another data quality problem mentioned in the discussion was incorrect data. The greatest challenge is very often to identify incorrect data. So far there are no formal procedures for this purpose. Once the erroneous data are identified then several options exist, like discard or correct the data. Another possibility is to use the incorrect data in the model without correcting them, but assessing the impact of this data and informing the user about the confidence of the prediction based on the available data. This sensitivity analysis can be done with ensemble techniques.

A similar problem to the previous one is inaccurate data. Naturally one can try to get more precise data, build more accurate sensors, etc. But always those measurements will have some measurement error. Therefore a better solution is to use methods that can deal internally with this uncertainty of data, and also make the final user aware of the magnitude of this uncertainty.

The final problem discussed concerned contradictory data. Current approaches usually suggest selecting one of the variables and discarding the other. Another option is to build models with different fragments of contradictory data and only later fuse or aggregate the output of the models.

## 5. Concluding remarks

Big data enables new opportunities for supply chain and logistic business. However data quality issues presents huge and costly problems for many companies. This report describes the results of the eSCF Awareness Workshop Handling Uncertainty for Data Quality Management - Challenges from Transport and Supply Chain Management, with purpose of raising awareness into data quality management issues. The discussion of the results from this workshop shows that many of the problems can be avoided or solved if a proper data management strategy is in place.